\documentclass[aps,prd,twocolumn,showpacs,nofootinbib]{revtex4-1}

\usepackage{amsmath}
\usepackage{amsfonts,amssymb}
\usepackage{multirow}
\usepackage[dvips]{graphicx}
\usepackage{float}
\usepackage{appendix}
\usepackage{color}
\usepackage{subfigure}
\usepackage[colorlinks]{hyperref}
\usepackage{ulem} 

\newcommand{\Ima}{{\rm Im}\,}

\def\beq{\begin{equation}}
\def\eeq{\end{equation}}
\def\bea{\begin{eqnarray}}
\def\eea{\end{eqnarray}}
\def\nn{\nonumber}

\def\chic1{\chi_{c1}}

\newcommand{\zbl}{Z_b(10610)}
\newcommand{\zbh}{Z_b(10650)}

\begin{document}

\title{General considerations on the nature of $Z_b(10610)$ and $Z_b(10650)$ from their pole positions } 
\author{Xian-Wei Kang $^1$}\email{xianwei.kang@um.es}
\author{Zhi-Hui Guo $^{2,3}$}\email{zhguo@mail.hebtu.edu.cn}
\author{J.~A.~Oller $^1$}\email{oller@um.es}
\affiliation{$^1$ Departamento de F{\'i}sica, Universidad de Murcia, E-30071 Murcia, Spain\\
$^2$ Department of Physics, Hebei Normal University, Shijiazhuang 050024, People’s Republic of China\\
$^3$ Helmholtz-Institut f\"ur Strahlen- und Kernphysik and \\
                        Bethe Center for Theoretical Physics,  Universit\"at Bonn, D-53115 Bonn, Germany}

\begin{abstract}
The nature of the bottomonium-like states $Z_b(10610)$ and $Z_b(10650)$ is studied by  calculating the $B^{(*)}\overline B^{*}$
 compositeness ($X$) in those resonances. We first consider uncoupled isovector $S$-wave scattering of 
$B^{(*)}\overline B^{*}$   within the framework of effective-range expansion (ERE). Expressions  for  the scattering length ($a$) and effective range ($r$) 
 are derived exclusively in terms of the masses and widths of the two $Z_b$ states.
  We then develop compositeness within ERE for the resonance case and deduce the expression $X=1/\sqrt{2r/a-1}$, which is 
then applied to the systems of interest. Finally, the actual compositeness parameters 
are calculated in terms of resonance pole positions and their experimental branching ratios 
into $B^{(*)}\overline{B}^*$ by using the method of Ref.~\cite{Oller1507}. 
 We find the values  $X=0.66\pm 0.11$  and $0.51\pm 0.10$  
  for the $Z_b(10610)$ and $Z_b(10650)$, respectively.  We also compare the ERE with Breit-Wigner and Flatt\'e parameterizations  
to discuss the applicability of the last two  ones for near-threshold resonances with explicit examples. 
\end{abstract}

\maketitle

\section{Introduction}

The discovery of exotic $XYZ$ mesons, specially those with hidden charm or bottom quarks, 
opens a new era in hadron physics spectroscopy. Since the observation of $X(3872)$~\cite{Choi:2003ue}, 
more than twenty $XYZ$ mesons are observed by several experimental collaborations in last decade~\cite{pdg}, for  recent reviews
 see Refs.~\cite{XLiu-overview,Olsen,Chen:2016qju}. 
The conventional quark potential model, which is successful in describing the heavy quarkonium below 
the open heavy-flavor threshold, typically cannot well accommodate these $XYZ$ states.  
Interestingly many of these exotic states share the common feature of lying nearby the thresholds 
 of pairs of open charm/bottom mesons. 
 In this respect, the effective range expansion (ERE) approach could  provide a proper framework to study
 the physics in the vicinity of thresholds~\cite{021015.1.bethe,021015.2.preston}.

Amongst the many observed $XYZ$ mesons, we focus on the two bottomonium-like states $\zbl$($Z_b$) and $\zbh$($Z_b'$) in the present work. 
They were measured by Belle Collaboration in the invariant mass spectra of $\pi^{\pm}\Upsilon(nS)$ ($n=1, 2, 3$) and $\pi^{\pm}h_b(mP)$ ($m=1, 2$), 
through the $\Upsilon(5S)$ decays with an additional charged pion~\cite{BelleZb}. 
All these five channels yield consistent values for the mass and width of the resonances \cite{BelleZb}, which are (in units of MeV) 
\begin{align}
\label{mwzbl}
M_{Z_b}=& 10607.2\pm2.0 \,,\Gamma_{Z_b}= 18.4 \pm 2.4 \,,\\ 
M_{Z_b'}=& 10652.2\pm1.5 \,,\Gamma_{Z_b'}= 11.5 \pm 2.2 \,.\nn
\end{align}
  Later on, the two $Z_b$ states were also confirmed in 
the $\Upsilon(5S)\to [B\overline{B}^{*}+ c.c.]\pi$ and $B^{*}\overline{B}^{*}\pi$ channels \cite{BelleZbconf2}.
  Due to the noticeable feature that $Z_b$ and $Z_b'$ are charged bottomonium-like states, 
they can not be conventional heavy quarkonium. 
 The neutral state $Z_b^0(10610)$ was also observed in $\Upsilon(5S)\to \Upsilon(nS)\pi^0\pi^0$  \cite{BelleZbconf3} with 
mass and width  compatible with Eq.~\eqref{mwzbl}.  
 The observed $Z_b^{(\prime)}$   have quantum numbers $I^G(J^P)=1^+(1^+)$ \cite{BelleZb,1403}, so that 
their electrically neutral isotopic states ($I_3=0$) should have $J^{PC}=1^{+-}$.
 Regarding this observation,  the shorthand notation $B\overline B^*$   actually indicates  the  $G=+1$ combination
 $B\overline B^* - \overline  B B^*$. 

These exciting experimental observations have strongly intrigued  theorists \cite{XLiu-overview,Olsen,Chen:2016qju},
 and the properties of the $Z_b^{(\prime)}$  have been investigated in  various theoretical approaches with the main aim of unveiling their nature. 
 There are also proposals that the two $Z_b$ resonances represent  kinematical effects at thresholds, e.g.  Refs. \cite{Bugg,Swanson}
 and \cite{XLiu2011}. The former two references interpret them as cusp effects, while the latter one advocates 
the energetic initial-single-pion emission mechanism to explain the experimental peak structures.  
However, Ref.~\cite{FKGuo.120216.6} argues that such  interpretations are not consistent. 
 This reference shows that in order to produce  peaks as pronounced and narrow as observed in experiment, nonperturbative interactions among the 
heavy mesons are necessary, which also give rise to the emergence of a nearby pole. Reference~\cite{FKGuo.120216.6} also stresses the importance 
of measuring the transition of the near-threshold structure to the associated continuum channel (this is  measured for the $Z_b^{(\prime)}$ in 
Ref.~\cite{BelleZbconf2}) to properly check the assumption about a purely kinematic origin of near-threshold enhancements.

 Several dynamical explanations have been  proposed to study these resonances,  
including mesonic molecular 
states~\cite{Voloshin,Cleven1,Cleven2,Nieves,Guo:2013sya,PingJL,ZhuSL,liu.120216.8,DongYB,Oset,mehen.120216.9,ohkoda.120216.10,PingJL}, 
compact tetraquark~\cite{Ali:2011ug,Braaten:2014qka}, quark-gluon hybrid~\cite{Braaten:2014qka}, 
hadro-quarkonium~\cite{Dubynskiy:2008mq,danillkin.120216.5},  etc
 (see Refs.~\cite{XLiu-overview,Olsen,Chen:2016qju} 
for a more complete account of theoretical methods and related works). 
 Typically the different models are characterized by assuming different constituents inside $Z_b$ and $Z_b'$ resonances. 
In the molecular picture the $Z_b^{(\prime)}$ contain pairs of heavy-light mesons (color-singlet clusters) that largely do not overlap 
within the molecule and keep basically their  identities with small virtual three-momenta involved in the making up of the state 
 \cite{Voloshin,Cleven1,Cleven2,Nieves,Guo:2013sya,PingJL,ZhuSL,liu.120216.8,DongYB,Oset,mehen.120216.9,ohkoda.120216.10}. 
 On the contrary, more tightly packed states where the heavy and light quarks overlap would be generic multiquark states. 
  For instance,  two colored diquark-antidiquark clusters are assumed to be the dominant components in Ref.~\cite{Ali:2011ug}. 
The quark-gluon hybrid approach advocates that the heavy quark-antiquark pair is embedded in the gluon and
 light-quark fields~\cite{Braaten:2014qka}. 
In the hadro-quarkonium model, the $Z_b$ states are described as systems with a compact color-singlet heavy-quarkonium 
core surrounded by light hadronic fields extending over larger distances~\cite{Dubynskiy:2008mq}. Tetraquark and open-heavy meson molecule  pictures 
were also examined within QCD sum rules \cite{HuangMQ,WangZG,navarra.120216.3,cui.120216.4,WangZG-tetraquark} 
and chiral quark models \cite{PingJL,entem.120216.2}.
 Of course, it is possible that the observed $Z_b^{(\prime)}$ states could be an admixture of several proposed mechanisms.

In the present work we do not study any of the previous models in further detail,
 but focus on the energy regions in the vicinity of the $B\overline B^*$ and $B^*\overline B^*$ thresholds.
 We proceed in terms of general considerations from scattering theory, and no specific dynamics will be assumed.  
First we employ the ERE and later the method derived in Ref.~\cite{Oller1507}. 
  Due to the fact that the central values of the masses of $Z_b$ and $Z_b'$ are only around 2--3~MeV above 
the $B\overline{B}^*$ and $B^{*}\overline{B}^{*}$ thresholds, respectively, 
we study the elastic isovector $S$-wave $B^{(*)}\overline{B}^{*}$ scattering that has the same quantum numbers as the 
resonances.  We employ the uncoupled ERE for such aim up to the resonance pole positions. However, the nominal 
$B^{(*)}\overline{B}^*$ three-momentum at the pole position is around $m_\pi$ (with $m_\pi$ the pion mass), which is beyond the strict 
radius of convergence of the ERE, that cannot exceed the one-pion exchange branch point at momentum $k=im_\pi/2$. 
 Nevertheless, there are serious indications that pion exchanges are perturbative for the $I^G(J^P)=1^+(1^{+})$  $B^{(*)}\overline{B}^*$ scattering 
in the low-energy region near threshold. In this respect, we have the estimates of Ref.~\cite{pavon.pwc.230216.1}, based on power counting in effective field theory,  
  which establish that the iteration of pion exchanges is suppressed by a large expansion scale $\Lambda$, 
 with $\Lambda \gg m_\pi$. This strong suppression  confirms the observation  of Ref.~\cite{Oset} that the exchange of 
light $q\bar{q}$ is  Okubo-Zweig-Izuka (OZI) 
 suppressed for the isovector $B^{(*)}\overline{B}^*$ scattering,  a suppression that is seen in specific calculations too \cite{Oset,XLiu2011,Ke.170316.1}.
 The same power counting study of Ref.~\cite{pavon.pwc.230216.1}  establishes  that coupled-channel effects
 should be suppressed as well, at least a next-to-leading effect.
 This is further reinforced by the experimental measurement of the large branching ratios ($Br$) of 
$Z_b^{(\prime)}$ into  $B^{(*)}\overline{B}^*$ \cite{BelleZbconf2}.

As a result, the  ERE study of the uncoupled isovector $S$-wave $B^{(*)}\overline{B}^*$ scattering seems to be a realistic  first approximation 
to the study of the $Z_b^{(\prime)}$ resonances. 
  Real values for scattering length ($a$) and effective range ($r$) are then required and we  fix them by reproducing the corresponding pole positions, 
with the masses and widths given in Eq.~\eqref{mwzbl}.\footnote{Note that, 
in coupled channel scattering with missing channels the parameters in the ERE could be complex. For example, we refer to the analysis of the $p\bar p$
 and $n\bar n$  in Ref.~\cite{KangHM1}.} Negative values for $a$ and $r$ are found between $-1$ and $-2$~fm.
  These natural values for the ERE parameters  indicate that the generation of the $Z_b^{(\prime)}$
 resonances does not require of any fine tuning, contrarily to e.g. $S$-wave nucleon-nucleon scattering and the deuteron or $\pi\Sigma_c$ scattering
 and the $\Lambda_c(2595)^+$ resonance \cite{guo16,hyodo.170216.1}. 
 We also derive a general interpretation of compositeness ($X$) for resonances in ERE that is applicable  as long as the resonance mass 
is above the corresponding threshold. If this condition is satisfied  we find that $X=1/\sqrt{2r/a-1}$, plus  contributions that would stem from higher orders in the ERE expansion. 
 Based on this discussion we derive the scaling with the heavy-quark mass for the mass and width of a resonance whose composition is saturated by the
 two-heavy-particle component. This occurs when the resonance width is  much larger than the difference between the resonance mass and the nearest threshold, a 
 pattern that is approximately realized in many $XYZ$ resonances (including the  $Z_b^{( \prime)}$ resonances). 
 Two simple potentials that can reproduce the values for $a$ and $r$ are discussed too. One of them contains only purely local interactions and the other 
is an attractive square-well potential. 
 Through the compositeness analysis, one can infer the inner structures of the two $Z_b$ states and we find that in the ERE case both resonances 
are dominated by the $B^{(*)}\overline{B}^*$ component, $76\%$ ($Z_b$) and $68\%$ ($Z_b'$), with an error of around $10\%$. 

In the last part of the work, we make use of the method  of Ref.~\cite{Oller1507} that is based on the expressions for the compositeness of a resonance 
 (with mass higher than the lightest threshold) and its width (including effects from the mass distribution of the resonance due to its finite width). 
Its applicability does not depend on the strength of nearby branch points from crossed-channel dynamics nor on the presence of
 other channels.\footnote{As long as the assumed Lorentzian mass distribution for the resonance is not strongly distorted. \label{foot.020316.1}}  
 Under the assumption that the partial decay widths of the $Z_b^{(\prime)}$ resonances into $B^{(*)}\overline{B}^*$ saturate the total widths we 
then reproduce  the results of the ERE study.
 However, we can use this other method to derive the actual compositeness coefficients by taking into account the experimentally 
measured branching ratios of the $Z_b^{(\prime)}$ into $B^{(*)}\overline{B}^*$ decay channels by the Belle Collaboration \cite{BelleZbconf2}. 
We then find  $X=66\%$ ($Z_b$) and $51\%$ ($Z_b'$), with around a $10\%$ of error,  for the $B\overline{B}^*$ and $B^*\overline{B}^*$ weights, 
respectively.

 Notice that in this  work the  compositeness coefficients 
  are purely determined by the masses and widths of $Z_b^{(\prime)}$  resonances. 
In this respect it is clearly different from the procedure of Ref.~\cite{ChenGY}, 
where the authors  perform  a combined analysis of  data from the decays $\Upsilon(5S)\to h_b(1P, 2P)\pi^+\pi^-$
 and $\Upsilon(5S)\to B^{(*)}\overline B^*\pi$ in effective field theory from where they
 analyze the $B^{(*)}\overline B^*$ compositeness of $Z_b^{(\prime)}$, with values compatible with  ours.

Finally,  taking into account that the applicability of the ERE is not affected by the $B^{(*)}\overline{B}^*$ threshold, 
 while this is the case for both   Breit-Wigner and Flatt\'e parameterizations, we then compare the amplitude squared from ERE 
with these other functions. In this way, we can gain some insight into the appropriateness of Breit-Wigner 
parameterizations to study the $Z_b^{(\prime)}$, as done by the Belle Collaboration \cite{BelleZb,BelleZbconf2,BelleZbconf3}. 
  We obtain that there is a marked cusp effect below threshold that can be well reproduced by a Flatt\'e 
parameterization, but  the Breit-Wigner function accurately reproduces 
$|t(E)|^2$ for $E>M_{\rm th}$ (including the maximum of the amplitude squared which fixes the resonance mass). As a result, 
we conclude that to apply  Breit-Wigner functions to study the $Z_b^{(\prime)}$ is not unrealistic, at least as a 
first approach. Of course,  sounder  parameterizations are clearly required to improve accuracy. 
 We also offer a variant of the Flatt\'e parameterization that exactly reproduces the amplitude squared calculated from the ERE.

The article is organized as follows. After this Introduction we develop the ERE study of uncoupled isovector $S$-wave $B^{(*)}\overline B^*$
 scattering around threshold in Sec.~\ref{sec:031115.1},
 where the resulting values for $a$, $r$ and the residues of the partial-wave amplitudes at the resonance pole positions are calculated. 
 There we also adapt to nonrelativistic kinematics the calculation of compositeness $X$ in Ref.~\cite{Oller1508}
  and  derive an algebraic expression  in terms of resonance mass and widths, Sec.~\ref{sec:160216.1}. 
 These results are then applied to the $Z_b^{(\prime)}$ resonances. In Sec.~\ref{sec:150216.1} we consider the limit $X\to 1$ and the
 evolution of the pole position with the heavy quark mass. Section~\ref{sec:290216.1} discusses two potentials 
 that reproduce the values of $a$ and $r$ found  for the $1^+(1^+)$ $S$-wave $B^{(*)}\overline{B}^*$ scattering.  
 In Sec.~\ref{sec:031115.2} we apply the method of Ref.~\cite{Oller1507} 
and   calculate  the final compositeness coefficients in $B^{(*)}\overline{B}^*$  of $Z_b^{(\prime)}$ by taking into account their experimental branching ratios. 
 Section~\ref{sec.170316.1} contains a discussion about the question of applicability of Breit-Wigner and Flatt\'e parameterizations 
to fit data in the study of these resonances. 
 A summary of the results and conclusions are given in Sec.~\ref{sec:conclusion} .

\section{Effective range study}
\label{sec:031115.1}

As explained in the Introduction we consider first the study of the uncoupled isovector $S$-wave $B^{(*)}\overline B^*$ scattering 
around  threshold, that has the same quantum number as the $Z_b^{(\prime)}$ resonances. 
 At first glance, the interest in this partial wave seems justified since the masses of $Z_b^{(\prime)}$ resonances are almost on top of the thresholds
 of $B^{(*)}\overline B^*$. Nonetheless, the nominal three-momenta of these particles in the pole positions corresponding to the resonances have  
a modulus of around $m_\pi$,  
so that a closer look is needed to justify the use of the ERE for the study of the resonances $Z_b^{(\prime)}$. Indeed the one-pion branch point is located at 
$k=i m_\pi/2$ and then, strictly, the radius of convergence of the ERE is smaller than the distance to the resonance pole positions. 
At this stage, we make use of advances in the literature where it is shown that pion exchanges in the isovector $B^{(*)}\overline{B}^*$ scattering  
 are expected to be clearly perturbative, since their iteration is suppressed by a large scale $\Lambda\gg m_\pi$ \cite{pavon.pwc.230216.1}.
 This outcome is based on derivations  from power counting in chiral effective field theory \cite{pavon.pwc.230216.1,fleming.240216.1,kaplan.240216.2}. 
 This power counting  establishes that coupled-channel effects are suppressed too, at least up to next-to-leading order. On the other hand, we also have   
the interesting observation of Ref.~\cite{Oset}  which found that one light $q\bar{q}$-meson  exchanges violate the OZI rule and are suppressed, 
 while stressing  the role of heavy-vector exchanges (that would correspond to contact interactions at low energy). 
 Similar results were also obtained in previous studies in the charmonium sector \cite{aceti.240216.3,Nieves}.   
 Then, we consider that for studying the $Z_b^{(\prime)}$ the use of the ERE up to their pole positions in
 the $1^+(1^{+-})$ $B^{(*)}\overline{B}^*$ uncoupled $S$-wave scattering seems a well suited first approximation. 

 The fact of disregarding  coupled channel effects 
 with other states that have  different particle content to $B^{(*)}\overline{B}^*$,  
 implies  that the total widths of the $Z_b^{(\prime)}$ resonances must be saturated by their partial decay widths into $B^{(*)}\overline{B}^*$.
 This is a strong conclusion from the previous considerations that indeed can be checked experimentally since these branching ratios  
 have been  measured  by Belle Collaboration in Ref.~\cite{BelleZbconf2},  where the following values are reported:
\begin{align}
\label{220216.3}
Br(Z_b(10610)^+)\to B\overline{B}^*)=&(86.0\pm 3.6)\,\%~,\\
Br(Z_b(10650)^+)\to B^*\overline{B}^*)=&(73.4\pm 3.6)\,\%~.\nn
\end{align}
We see that both $Br$'s are rather large, which supports our way of proceeding.

The ERE gives rise to nonrelativistic $B^{(*)}\overline{B}^*$ partial-wave amplitudes
 that  have exclusively right-hand cut or unitarity cut, without  crossed-channel cuts. 
 The general expression for a partial wave when crossed-channel cuts are absent 
 is derived in Ref.~\cite{300915.2.ndorg} by making use of the $N/D$ method \cite{051015.1.chew}, and it can be expressed as 
\begin{align}
\label{271015.3}
t(E)=&\left[\sum_i \frac{g_i}{E-M_{i,\rm CDD}}-i k \right]^{-1}~.
\end{align}
Regarding the kinematical variables used in this equation, $E$ is the center-of-mass (CM) energy of the system 
and $k$ is the CM on-shell three-momentum.  The nonrelativistic relation between $E$ and $k$ is valid for the present case 
and it reads 
\begin{equation}
\label{271015.4k}
k=\sqrt{2\mu(E-M_{\rm th})}\,,
\end{equation}
where $\mu=m_1 m_2/(m_1+m_2)$ is the reduced mass for the system with masses $m_1$ and $m_2$ and 
$M_{\rm th}=m_1+m_2$ stands for the threshold.

The dynamical content of Eq.~\eqref{271015.3} is driven by the sum over the so-called Castillejo-Dalitz-Dyson (CDD) poles (each of them corresponds  to a zero of $t(E)$ at $E=M_{i,\rm CDD}$), so that the $i_{\rm{th}}$ CDD pole is given in terms of its residue $g_i$ and mass $M_{i,\rm CDD}$. 
The expansion in powers of $k^2$ of the CDD poles  around $k=0$ is equivalent to the ERE,  but  notice that 
this expansion would be valid until  the position of the nearest CDD pole to threshold.
 Up to including ${\cal O}(k^2)$ in the expansion of $k\cot\delta_0$ (with $\delta_0$
 the isovector $S$-wave $B^{(*)}\overline B^*$ phase shifts) 
the ERE of the $S$-wave amplitude can be written as
\begin{equation} 
\label{271015.1}
t(E)=\frac{1}{-\frac{1}{a}+\frac{1}{2}r\,k^2-i\,k}~.
\end{equation}

Here we have employed the same normalization as in Eq.~\eqref{271015.3}, such that along the physical axis one has the 
unitarity condition
\begin{align}
\label{271015.4}
{\rm{Im}}\, t(E)^{-1}=-k\leq 0~.
\end{align}
However, the convergent range of the ERE would be severely restricted if any of the CDD poles 
in Eq.~\eqref{271015.3} had a mass very close to $M_{\rm{th}}$.
 This could be the case if the $Z_b^{(\prime)}$ had  important components other than the $B^{(*)}\overline B^*$ ones, e.g. other heavier 
channels or corresponding to more elementary QCD degrees of freedom 
in terms of compact quark-gluon states  ~\cite{WangZG-tetraquark,Ali:2011ug,Braaten:2014qka,Dubynskiy:2008mq}. 
A distinctive feature of this situation can be recognized by considering the contribution of this CDD pole to $a$ and $r$, which reads 
\begin{align}
\label{281015.5}
\delta a=&- \frac{M_{\rm{th}}-M_{i,\rm CDD}}{ g_i }~,\\
\label{281015.5b}
\delta r=&-\frac{ g_i }{\mu (M_{\rm{th}}-M_{i,\rm CDD})^2}~.
\end{align}
As a result one should expect that if $M_{i,\rm CDD}\simeq M_{\rm{th}}$ a large absolute value of $r$ would arise,
 because the denominator in Eq.~\eqref{281015.5b} would involve the square of a small quantity (of the order of a kinetic energy, which also compensates 
for the appearance of the factor $\mu$ in this denominator). 
In this case the effective range $r$ would be  much larger than a standard value from potential scattering  \cite{021015.1.bethe,021015.2.preston}, 
which typically would be similar to the  natural range of strong interactions, 
$m_\pi^{-1} \sim \Lambda_{\rm QCD}^{-1}\sim 1$~fm, with $\Lambda_{\rm QCD}$ the typical QCD non-perturbative scale \cite{pdg}. 
  Then, the new small energy scale $|M_{i,\rm CDD}-M_{\rm{th}}|$ could totally spoil the application of the ERE analysis to the $Z_b^{(\prime)}$ states around 
the $B^{(*)}\overline B^*$ thresholds. 
 We refer to Ref.~\cite{guo16} for a devoted discussion of the situation with $M_{\rm CDD} \simeq M_{\rm th}$ for the case of the resonance 
 $\Lambda_c(2595)^+$ almost on top of the thresholds of the channels $\pi^0\Sigma_c^+$, $\pi^+ \Sigma^0$ and $\pi^-\Sigma^{++}$. 

Taking into account the previous warning, we apply first the ERE for $t(E)$ at 
the pole position $E=E_R$ of  the resonance, with $E_R$ given by 
\begin{align}
\label{010316.1}
E_R=M_R-i\frac{\Gamma_R}{2}~,
\end{align}
 where  $M_R$ and $\Gamma_R$ are its mass and width, in order.\footnote{We have checked within the present ERE study that 
the resulting $t(E)$ above threshold is very well reproduced by a standard Breit-Wigner parameterization. So that  the identification in 
Eq.~\eqref{010316.1} is consistent here.}
  We denote by $t_{II}(E)$ the partial-wave amplitude $t(E)$ in the 2nd Riemann sheet (RS), where the resonance pole lies. 
Its explicit form reads  
\begin{equation}
\label{271015.2}
t_{II}(E)=\frac{1}{-\frac{1}{a}+\frac{1}{2}r\,k^2+i\,k}~.
\end{equation}
Notice  the change of sign in front of $k$, compared to Eq.~\eqref{271015.1}, with $k$ given by Eq.~\eqref{271015.4k} and calculated 
such that ${\rm Im}k>0$ (1st RS). We denote by $k_R$ the momentum at the resonance pole position, 
\begin{align}
\label{150216.1}
k_R=\sqrt{2\mu(E_R-M_{\rm{th}})}~,
\end{align}
 and write it as 
\begin{align}
\label{150216.2}
k_R=k_r+i\,k_i~,~k_i>0~.
\end{align}
Let us derive a more explicit expression for $k_R$. For that we introduce the angle $\phi$ ($0\leq \phi \leq \pi/4$) defined as
\begin{align}
 \label{170216.2}
\tan 2\phi=\frac{\Gamma_R}{2|M_R-M_{\rm th}|}~, 
\end{align}
 and in terms of it $k_R$ reads  
\begin{align}
\label{170216.1}
\rm{i)\,} & M_R-M_{\rm th}>0~,\\
k_R=&\sqrt{2\mu|M_R-M_{\rm th}-i\frac{\Gamma_R}{2}|}\,\exp[i(\pi-\phi)]\nn\\
=&\sqrt{2\mu|M_R-M_{\rm th}-i\frac{\Gamma_R}{2}|}\,(-\cos \phi+i \sin \phi)~.\nn\\
\label{190216.1}
\rm{ii)\,}& M_R-M_{\rm th}<0~,\\
k_R=&\sqrt{2\mu|M_R-M_{\rm th}-i\frac{\Gamma_R}{2}|}\,\exp[i(\frac{\pi}{2}+\phi)]\nn\\
=&\sqrt{2\mu|M_R-M_{\rm th}-i\frac{\Gamma_R}{2}|}\,(- \sin \phi+i\cos \phi)~.\nn
\end{align}
   The parameters $a$ and $r$ in the partial-wave amplitude $t(E)$ are fixed so as to reproduce the mass and width of the resonance,
 and from these values we can further discern whether there is  an indication of a nearby CDD pole or just the opposite,
 i.e., a situation corresponding to a pure $S$-wave potential scattering problem. 
 We  find $a$ and $r$ by requiring that $t_{II}(E)^{-1}=0$ at $E=E_R$, namely,
\begin{align}
\label{170216.3}
0=&-\frac{1}{a}+\frac{1}{2}r\,k_R^2+i\,k_R\\
=&-\frac{1}{a}+\frac{1}{2}r\,(k_r^2-k_i^2+2i k_r k_i)+i k_r-k_i~.\nn
\end{align}
Requiring the vanishing of both the real and imaginary parts of the previous equation, the following expressions result for $k_r\neq 0$ 
\begin{align}
\label{281015.3b}
a=&-\frac{2k_i}{|k_R|^2}~,\\
\label{281015.3}
r=&-\frac{1}{k_i}~.
\end{align}
Notice that given $k_r$ and $k_i$ one always finds $a$ and $r$ from Eqs.~\eqref{281015.3b} and \eqref{281015.3}, respectively. 
For the opposite case, that is, once  $a$ and $r$ are known, there is a resonance only if $a<0$, $r<0$ and $a/2>r$,
 as it is clear from Eqs.~(\ref{281015.3b},\ref{281015.3}). The case $k_r=0$ corresponds to a virtual state, $M_R<M_{\rm th}$ and $\Gamma_R=0$,  
 and only one equation results then, $1/a+r k_i^2 /2+k_i=0$.

Around the pole position we expand the denominator of $t_{II}(E)$ up to first order in $k-k_R$ and, by taking into account the resonance pole condition,  
 cf. Eq.~\eqref{170216.3}, the expression for $t_{II}(E)$ becomes
\begin{align}
\label{160216.5}
t_{II}(k)=\frac{1}{(rk_R+i)(k-k_R)}+\ldots=\frac{-k_i/k_r}{k-k_R}+\ldots
\end{align}
where we have used that $r=-1/k_i$, Eq.~\eqref{281015.3}, and 
the ellipsis indicate higher order terms in the expansion in powers of $k-k_R$. From this equation we directly find 
 the residue in the variable $k$,
\begin{align}
\label{160216.6}
\gamma_k^2=&-\frac{k_i}{k_r}>0~.
\end{align}
Notice that $k_r<0$, as it is clear from Eq.~\eqref{170216.1}.

The residue of $t_{II}(E)$ in the standard Mandelstam variable $s=E^2$ is denoted by $-\gamma^2$, 
\begin{align}
\label{281015.2}
t_{II}(E)\xrightarrow[E \to E_R]{} -\frac{\gamma^2}{s-E_R^2}~.
\end{align}
 Then, the  relation between $\gamma^2$ and $\gamma_k^2$ follows straightforwardly 
\begin{align}
\label{190216.9}
\gamma_k^2=-\gamma^2 \left.\frac{dk}{ds}\right|_{k_R}=-\frac{\mu \gamma^2}{2E_R k_R}~.
\end{align}

When solving $1/t_{II}(E_R)=0$ for $a$ and $r$, we also take into account the uncertainties in the pole positions 
of $Z_b$ and $Z_b'$ in Eq.~\eqref{mwzbl}. The numerical results are then given in Table~\ref{tabar}.  
 In order to implement the error estimate, we discretize the pole mass and width at several points within around the one and a half $\sigma$ region from the central values, 
so that a data grid results. For each of the points in the grid we calculate the corresponding $a$ and $r$,  cf. Eqs.~(\ref{281015.3b},\ref{281015.3}),
and the central values are given by the respective mean values and the errors by the square root of the variances. 
The procedure is of course stable by increasing the number of points in the grid and, e.g.,
convergence is already found when  nine points in equal step for the mass and width are taken. 
The calculation of other quantities that stem from the knowledge of $a$ and $r$ will also follow this procedure. 

\begin{table}[htbp]
\begin{center}
\begin{tabular*}{\linewidth}{@{\extracolsep{\fill}}lrr}  
\hline \hline
 {} &$Z_b(10610)$   &$Z_b(10650)$ \\
\hline
$a\,(\rm{fm})$ & $-1.03 \pm 0.17$  & $-1.18 \pm 0.26$\\
$r\,(\rm{fm})$ & $-1.49 \pm 0.20$ & $-2.03 \pm 0.38$ \\
$X=\gamma_k^2$ & $0.75\pm 0.15$  & $0.67\pm 0.16$\\
$g^2\,(\rm{GeV}^2)$ & $362 \pm 71$  & $ 263 \pm 63$\\
\hline\hline
\end{tabular*}
\caption{From top to bottom and left to right,
 we give the scattering lengths ($a$) and effective ranges ($r$) of the $B\overline B^{*}$ ($Z_b$) and $B^{(* )}\overline B^{*}$ ($Z_b'$) systems, in order. 
In the last two lines  compositeness ($X$), which is equal to $\gamma_k^2$,  and couplings squared $g^2$  for the $Z_b^{(\prime)}$ resonances are collected. }
\label{tabar}
\end{center}
\end{table}

As it is clearly seen from Table~\ref{tabar}, the values for $r$ correspond to the typical range of strong interactions, as they are of the order of $1/\Lambda_{\rm QCD}$. 
 Because of this $r$ behaves as  expected for potential scattering  \cite{021015.1.bethe,021015.2.preston}.  
In particular, it excludes the possibility of having a nearby CDD pole around threshold because, as discussed above, it would give rise 
to large contributions in absolute value to $r$ \cite{guo16}. This fact,  
is also an indication that the $Z_b^{(\prime)}$  can be understood to large extent as $B^{(*)}\overline B^{*}$ resonances (in the case of uncoupled  scattering).

\subsection{Compositeness for a resonance within ERE}
\label{sec:160216.1}

In order to further quantify the statement on the nature of the $Z_b^{(\prime)}$ as $B^{(*)}\overline B^{*}$ composite resonances 
 we apply here  the theory developed in Ref.~\cite{Oller1508}, that  allows 
a probabilistic interpretation of the compositeness relation~\cite{Weinberg:1962hj,Baru:2003qq,ref.230715.4,ref.230715.5,ref.230715.6,Agadjanov:2014ana} 
for those resonances such that $\sqrt{{\rm Re}E^2_R}$ is larger than the lightest threshold.  
We adapt the procedure of Ref.~\cite{Oller1508} to nonrelativistic kinematics,  that is also the one employed in the ERE,
 cf. Eqs.~\eqref{271015.1} and \eqref{271015.4}.

  We can follow analogous steps for  the derivation of the criterion of  applicability for
 the compositeness relation as done in Ref.~\cite{Oller1508}  but now in the variable $E$ instead of $s$. 
 This change of variable is motivated by the fact that we are considering a nonrelativistic system.  
  The Laurent series for $t_{II}(E)$ around $E=E_R$ in powers of $E-E_R$ has $|E_R-M_{\rm th}|$ as radius of convergence because of 
 the branch point at threshold, cf. Eq.~\eqref{271015.4k}.  As a result, as long as $M_R>M_{\rm th}$ this Laurent series always embraces a portion of the physical real axis 
and the probabilistic interpretation of $X$ as the weight  in the composition of the resonance then follows \cite{Oller1508}. Notice that within the relativistic formalism 
of the latter reference the criterion for the applicability of this result is more restrictive because whenever $\sqrt{{\rm Re}E^2_R}>M_{\rm th}$ 
then  $M_R>M_{\rm th}$ is fulfilled and 
no contradiction arises between our present criterion and that in Ref.~\cite{Oller1508}. 
At the practical level in our present application to the $Z_b^{(\prime)}$ resonances there is no difference between both 
criteria because ${\rm Re}E^2_R=M_R^2-\frac{\Gamma_R^2}{4}$ and $(\Gamma_R/2 M_R)^2\sim 10^{-6}$.\footnote{For the compositeness criterion an expansion 
in energy not in momentum is employed because one is extrapolating in the resonance mass from the narrow resonance case to  the inner complex plane.}

The nonrelativistic reduction of the unitarity loop function $G(E)$, with normalization 
\begin{align}
\label{190216.3}
\Ima  G(E)=- k~,~k>0
\end{align}
according to the unitarity condition Eq.~\eqref{271015.4}, is given by the integral representation ($\Ima E\neq 0$)
\begin{align}
\label{010316.2}
G(E)=&-\frac{E-M_{\rm th}}{\pi}\int_{M_{\rm th}}^\infty dE'\frac{\sqrt{2\mu(E'-M_{\rm th})}}{(E'-M_{\rm th})(E'-E)} \\
+&G(M_{\rm th})~.\nn
\end{align}
 One subtraction has been taken at threshold to end with a convergent integration,
 with the subtraction constant $G(M_{\rm th})\in \mathbb{R}$. By rewriting $E'=M_{\rm th}+q^2/2\mu$ in Eq.~\eqref{010316.2} the latter becomes 
\begin{align}
\label{190216.4}
G(E)=&-\frac{k^2}{\pi}\int_0^\infty dq^2\frac{1}{q(q^2-k^2)}+G(M_{\rm th})\\
=&-ik+G(M_{\rm th})~.\nn
\end{align}
It need not be reiterated that $G(E)$ in the 1st RS is calculated with $\Ima k\geq 0$, while in the 2nd RS $\Ima k \leq 0$. In this 
latter case we denote the function $G(E)$ as $G_{II}(E)$. To avoid possible confusion with respect to the RS in which $k$ is calculated 
we write our formulas such that $k$ is calculated always in the 1st RS. In this from, $G_{II}(E)$ reads 
\begin{align}
\label{160216.7}
G_{II}(E)=&i k+G(M_{\rm th})~,~{\rm Im}k>0~.
\end{align}
From this equation we also obtain
\begin{align}
\label{190216.5}
\frac{d G_{II}(E)}{d k}&= i~.
\end{align}

Then, for $M_R\geq M_{\rm th}$ we have, according to procedure of Ref.~\cite{Oller1508}, the following expression for the 
compositeness $X$, 
\begin{align}
\label{190216.6}
X=\left|\gamma^2\frac{d G(E_R)}{d s}\right|=\left|\gamma^2 \frac{d k}{d s}\frac{d G(E_R)}{d k}\right|^2=|\gamma_k|^2~,
\end{align}
in virtue of the relation between the residues $\gamma^2$ and $\gamma_k^2$, cf.  Eq.~\eqref{190216.9}. 
Notice that $X$ is independent of the subtraction constant because it disappears in $dG(E_R)/dk$.

Now, if we restrict ourselves to the ERE of $t_{II}(E)$ up to including ${\cal O}(k^2)$, Eq.~\eqref{271015.2}, the residue  
  $ \gamma_k^2$ is given by Eq.~\eqref{160216.6}, and then $X$ reads
\begin{align}
\label{190216.11}
X=- \frac{k_i}{k_r}=\tan\phi\leq 1~,
\end{align}
because $\phi\in [0,\pi/4]$, cf. Eq.~\eqref{170216.2}.
 According to Eqs.~\eqref{170216.1} and \eqref{190216.1}, the upper bound in the previous equation holds for $M_R\geq M_{\rm th}$,  
which shows the importance of fulfilling the criterion for compositeness so as to ascribe this meaning to $X$. 
  Furthermore, the condition  $M_R\geq M_{\rm th}$ deduced in the nonrelativistic discussion also provides a
 concrete example of the method of Ref.~\cite{Oller1508}. 
 The elementariness $Z$ is defined as 
\begin{align}
\label{190216.10}
Z=&1-X~,
\end{align}
and $0\leq Z \leq 1$ because of Eq.~\eqref{190216.11}.  
 We can also express $X$ directly in terms of the observable quantities $a$ and $r$ since 
\begin{align}
\label{190216.2}
X=-\frac{k_i}{k_r}=\left( \frac{2r}{a}-1\right)^{-\frac{1}{2}}~,
\end{align}
as follows from Eqs.~\eqref{281015.3b} and \eqref{281015.3}.

 The saturation of the equality $X=1$ occurs  only for  $M_R=M_{\rm th}$ and $\Gamma_R\neq 0$ ($\phi=\pi/4$) or, in other terms,  
 when $a=r$, according to  Eq.~\eqref{190216.2}. 
 In our present study this limit situation would correspond to $Z_b^{(\prime)}$ resonances  composed purely of two heavy mesons. 
 
 Substituting Eq.~\eqref{170216.2} into Eq.~\eqref{190216.11} and performing the expansion in powers
 of $(M_R- M_{\rm th})/\Gamma_R$,  another appealing way to express $X$ is  
\begin{align}
\label{xnew}
 X =& -\frac{2(M_R- M_{\rm th})}{\Gamma_R}+ \sqrt{1+\left[\frac{2(M_R- M_{\rm th})}{\Gamma_R}\right]^2} \\ \nn
=& 1-\frac{2(M_R- M_{\rm th})}{\Gamma_R} +2\left[\frac{(M_R- M_{\rm th})}{\Gamma_R}\right]^2 + \cdots \,,
\end{align}
where only the first three terms in the expansion are kept in the last line. 
 This expression is explicitly given in terms of    mass and width of the resonance,  
while  in  Refs.~\cite{Weinberg:1962hj,Baru:2003qq,ref.230715.4,ref.230715.5,ref.230715.6,Agadjanov:2014ana} the  coupling strengths
 are usually needed to obtain  $X$. 

The experimental values of $M_R$ for the $Z_b^{(\prime)}$ resonances fulfill the compositeness criterion $M_R\geq M_{\rm th}$, namely,  
\begin{align}
\label{160216.2}
M_{Z_b^+}-\big[M_{B^+}+ M_{\overline{B}^{*0}}\big]=&(3.1\pm 2.0)~\rm{MeV},\nn\\
M_{{Z'_b}^+}- \big[M_{{B^*}^+}+ M_{\overline{B}^{*0}}\big]=&(2.7\pm 1.6)~\rm{MeV},
\end{align}
where we have employed the latest values for the masses from the PDG \cite{pdg,BelleZb}.
  Thus, we can calculate the compositeness coefficient $X$ from Eq.~\eqref{190216.11} within the present ERE of uncoupled isovector $S$-wave  
$B^{(*)}\overline{B}^*$ scattering  and the values are given in the 4th row of Table~\ref{tabar}.
  The approximated expression in the last line of Eq.~\eqref{xnew}
 leads to $X=0.73\pm0.15$ for $Z_b$ and $0.66\pm0.15$ for $Z'_b$, which are in good agreement with the exact
 numbers in Table~\ref{tabar}.  
These results  indicate a large and dominant component of  $B^{(*)}\overline B^{*}$ for the resonances $Z_b^{(\prime)}$, though 
still other components are not negligible.
A sharper value of $X$ would require to improve the precision in the experimental determination of the mass and width of the resonance. 

\subsection{Scaling of the width  in heavy-meson composite resonances}
\label{sec:150216.1}

For the $Z_b^{(\prime)}$ resonances one has that the (small) width is significantly larger than the difference between the resonance mass and 
the close $B^{(*)}\overline{B}^*$ heavy-meson threshold, cf. Eqs.~\eqref{mwzbl} and \eqref{160216.2}.  
This is also a typical situation for many of the reported $XYZ$ resonances. 
If this is the case and the ERE can be applied, as argued above this is expected to be a good (first) approximation for the $Z_b^{(\prime)}$, 
 one can derive a simple way a scaling law for the width  of the resonance. 
 Let us recall that according to the discussion in Sec.~\ref{sec:160216.1} 
 the limit $2(M_R-M_{\rm th})/\Gamma_R\to 0$ correspond to a resonance made purely by the two heavy-mesons ($X\to 1$).

The resonant momentum $k_R$ in the limit $\Gamma_R/2\gg |M_R-M_{\rm th}|$ becomes in good approximation,  
\begin{align}
\label{150216.3}
k_R&\approx \sqrt{-i \mu \Gamma_R}
=-e^{-i\frac{\pi}{4}}\sqrt{\mu\Gamma_R}~,
\end{align}
 as follows by applying Eq.~\eqref{170216.1} with $\phi=\pi/4$ and 
neglecting $|M_R-M_{\rm th}|$ in front of  $\Gamma_R/2$.   
  As a result 
\begin{align}
\label{150216.4}
-k_r\approx k_i\approx \sqrt{\frac{\mu \Gamma_R}{2}}~. 
\end{align}
 This result implies because of Eq.~\eqref{281015.3} that 
\begin{align}
\label{150216.5}
r\approx - \sqrt{\frac{2}{\mu\Gamma_R}}~.
\end{align}
 We further require that $r$ has a  natural size for strong interactions, 
so that a CDD pole near threshold, which would give rise to a large contribution to $r$ for $M_{\rm CDD}\to M_{\rm th}$,
 cf. Eq.~\eqref{281015.5},  is excluded.  
 Then, with $r\sim- \frac{1}{\Lambda_{\rm QCD}}$ 
 we find from Eq.~\eqref{150216.5} an order of magnitude estimate for  $\Gamma_R$, 
\begin{align}
\label{150216.6}
\Gamma_R \sim \frac{2\Lambda_{\rm QCD}^2}{\mu}
\end{align} 
Since we assume that $\Gamma_R/2\gg|M_R-M_{\rm th}|$ one has the consistency requirement that
\begin{align}
\label{150216.7}
\frac{\Lambda_{\rm QCD}^2}{\mu}\gg|M_R-M_{\rm th}|.
\end{align}
Eqs.~\eqref{150216.6} and \eqref{150216.7} imply that as the heavy-quark mass increases 
the resonances composed mainly by the heavy-flavor mesons should become narrower and
 basically sit on top of threshold in the energy plane.
 However, let us notice that $k_R$ is stable because of the product of $\mu$ and $\Gamma_R$,
\begin{align}
\label{160216.1}
k_R\approx \sqrt{-i\mu\Gamma_R}\sim -e^{-i \frac{\pi}{4}}\sqrt{2}  \Lambda_{QCD}~.
\end{align}

The estimate of the width for the $B^{(*)}\overline{B}^*$  systems  from Eq.~\eqref{150216.6} gives  
$\Gamma_R\sim 30$~MeV. Here  we have taken $\Lambda_{\rm QCD}\simeq 200$~MeV, because the momentum transfer 
is small and this corresponds to the first few low-mass quark flavors. The estimated value for $\Gamma_R$ agrees with the experimental 
 widths of the $Z_b^ {(\prime)}$, Eq.~\eqref{mwzbl}, within  a  factor of around 2. The consistency check of Eq.~\eqref{150216.7} is 
well fulfilled because the differences between resonance masses and thresholds, cf. Eq.~\eqref{160216.2}, 
are considerably smaller than $\Lambda_{\rm QCD}^2/\mu\sim 15$~MeV.

\subsection{Two simple models as examples} 
\label{sec:290216.1}

Here we present two potentials that reproduce exactly the values of $a$ and $r$ deduced above from the ERE 
application to the study of the pole positions of the $Z_b^{(\prime)}$ resonances.

As a first example, we consider a  pure contact theory with an $S$-wave potential of the form
\begin{align}
\label{290216.1}
v(p,k)=C_0+C_2(p^2+k^2)~,
\end{align}
with $p$ and $k$ three-momenta. In terms of this potential one can reproduce any negative values for $a$ and $r$.
 This can be explicitly shown by solving the corresponding  Lippmann-Schwinger equation
\begin{align}
\label{290216.2}
t(p,k;E)=&v(p,k)+\frac{2}{\pi}\int_0^\infty dq\frac{q^2 v(p,q)t(q,k;E)}{q^2-k^2-i 0^+}~,
\end{align}
such that $E=k^2/2\mu$ and the on-shell $T$-matrix element, $t(k,k;k^2/2\mu)$  corresponds to $t(E)$, already introduced. 
One can easily solve Eq.~\eqref{290216.2} by proposing a solution of the form
\begin{align}
\label{290216.3}
t(p,k;k^2/2\mu)=&t_0(k)+t_2(k) (k^2+p^2)~.
\end{align} 
The integrations in the variable $q^2$ can be done with a three-momentum cut-off $\Lambda$.
 For a given $\Lambda$ one  then solves $C_0$ and $C_2$ such that the values for $a$ and $r$ are reproduced. 
 This makes that both $C_0$ and $C_2$ become  function of $\Lambda$. 
Real solutions $C_0(\Lambda)$ and $C_2(\Lambda)$ exist whenever $r<0$, e.g. we have for $C_2(\Lambda)$,
\begin{align}
\label{290216.4}
 C_2 \Lambda^3=-\frac{3\pi}{2}\left\{1\pm \frac{\sqrt{3}(\pi-2\alpha)}{\sqrt{3(2\alpha-\pi)^2+4\alpha^2-\alpha^2 \rho \pi}}\right\}~,
\end{align}
where $\alpha=a \Lambda$ and $\rho=r  \Lambda$. 
 The crucial property to end with real counterterms (as required for the potential in Eq.~\eqref{290216.1} to be Hermitian) is to demand that 
the radicand in the previous equation be positive,
\begin{align}
\label{290216.5}
3(2\alpha-\pi)^2+4\alpha^2-\alpha^2 \rho \pi\geq 0~,
\end{align}
 which is always the case if $r<0$ $(\rho<0$). 
 In the limit $\Lambda\to \infty$ one obtains exactly the ERE approximation for $t(E)$ of Eq.~\eqref{271015.1}, 
\begin{align}
\label{290216.6}
t(E)=\frac{1}{-\frac{1}{a}+\frac{1}{2}r k^2-ik}+{\cal O}(\Lambda^{-1})~.
\end{align}
 Ref.~\cite{kolck.290216.1} finds that   a renormalized  effective field theory with only 
contact interactions gives rise to an ERE (although the reverse is not always true \cite{phillips.290216.2}, cf. restriction in Eq.~\eqref{290216.5}). 
 Thus, a contact interaction theory with two couplings can exactly reproduce the results obtained before 
in the ERE study, for which $a$ and $r$ are negative.

The second example corresponds to a simple attractive square-well potential of radius $R$ ,
\begin{align}
\label{290216.7}
v(r)=-V_0 \theta(R-r)~,
\end{align}
with $V_0>0$. 
The $S$-wave amplitude can be easily found, e.g. by solving the corresponding Schr\"odinger equation
\begin{align}
\left \{
\frac{d^2}{dr^2}
+2\mu V_0 \theta(R-r)+k^2
\right\}
u(r)=0~,
\end{align}
with $u(r)$ the reduced wave function. This is explicitly solved e.g. in Ref.~\cite{stoof.290216.3}. 
The resulting $S$ matrix is 
\begin{align}
\label{290216.8}
S(E)=&e^{2i\delta_0(E)}\\
=&e^{-2ikR}
\frac{e^{ik' R}(k+k')+e^{-ik'R}(k'-k)}{e^{-ik'R}(k+k')+e^{ik'R}(k'-k)}~,\nn
\end{align}
where
\begin{align}
\label{290216.9}
k'=&\sqrt{k^2+2\mu V_0}~.
\end{align}
 The ERE is obtained from the expansion in 
powers of $k^2$ of   
\begin{align}
\label{290216.10}
k\cot\delta_0=&k\cot \left[-k R+{\rm arctan}\left\{\frac{k}{k'}\tan \left(k'R  \right)\right\}
\right]~.
\end{align}
The  corresponding expressions for $a$ and $r$ are
\begin{align}
\label{290216.11}
a=&R\left(1-\frac{\tan \gamma}{\gamma}\right)~,\\
\label{290216.11b}
r=&R\left(1-\frac{1}{\gamma^2 x}-\frac{1}{3x^2}\right)~,
\end{align} 
with the dimensionless variables 
\begin{align}
\label{290216.12}
\gamma=&R\sqrt{2\mu V_0}~,\\
x=&a/R~.\nn
\end{align}
 If we consider the application of this toy model to the isovector $S$-wave $ B^{(*)} \overline{B}^*$ scattering 
 one expects that the value of $\gamma^2\gg 1$ by taking the simple dimensional estimates $R\simeq 1/\Lambda_{QCD}$ and 
 $V_0\simeq \Lambda_{QCD}$, together with 
 $\mu\simeq M_B/2\simeq 2.6$~GeV, where $M_B$ is the mass of the $B$ meson.\footnote{The departure of the estimate $r\sim R$ (used in Sec.~\ref{sec:150216.1}) 
from Eq.~\eqref{290216.11b} with large $\gamma$  would require  $x\to 0$, which is 
far from our case since we have  $x\sim -1$, cf. Table~\ref{tabar}. Indeed for $x=0$ the 
ERE breaks down because then $k\cot \delta(0)=\infty$.}  Then 
\begin{align}
\label{290216.13}
\gamma^2 \sim  \frac{M_B}{\Lambda_{ \rm QCD}}  \sim 30~.
\end{align}
As a result, in order to end with a negative value for $a$ from Eq.~\eqref{290216.11} (so that $\tan\gamma/\gamma>1$), it is necessary that 
\begin{align}
\label{290216.14}
\gamma\to (n+\frac{1}{2})\pi~,
\end{align}
with $n\sim 1,\, 2$ to match the estimate in Eq.~\eqref{290216.13}. 
 
The parameter $x$ has a well-defined limit for $\gamma^2\to \infty$. 
 To obtain this conclusion we first notice that $x$ can be fixed by the ratio  $r/a$, since it results  from Eqs.~(\ref{290216.11},\ref{290216.11b}) that 
\begin{align}
\label{300216.1b}
\frac{r}{a}3x^3-3x^2+\frac{3x}{\gamma^2}-1=0~.
\end{align}
This equation  for $\gamma^2\gg 1 $ becomes independent of this parameter, and one has the following asymptotic equation
\begin{align}
\label{300216.2}
\frac{r}{a}3x_\infty^3-3x_\infty^2-1=0~.
\end{align}

Next, it is easy to show  that the difference between the actual 
 value of $\gamma$ and $(n+\frac{1}{2})\pi$ should scale as $1/\sqrt{\mu}$.
Calling this difference $\varepsilon$ we would have from Eq.~\eqref{290216.11} that
\begin{align}
\label{290216.15}
\frac{\tan\gamma}{\gamma}=1-x\to 1-x_\infty~,
\end{align}
which is a fixed number (around 1.5  by explicitly solving Eq.~\eqref{300216.2} with the values for $a$ and $r$ given in Table~\ref{tabar}.) Then, 
\begin{align}
\frac{\tan\gamma}{\gamma}=\frac{\tan\big[(n+1/2)\pi+\varepsilon \big]}{\gamma}\sim -\frac{1}{\varepsilon \gamma}=1-x_\infty~.
\end{align}
In this way we conclude that 
\begin{align}
 \label{290216.16}
\varepsilon \sim \frac{1}{\gamma}\sim \left(\frac{\Lambda_{\rm QCD}}{\mu}\right)^{\frac{1}{2}}~.
\end{align}
However,  $k_R$ corresponding to the resonance poles of $S(E)$
 does not run with $\mu$. This is clear if we consider the equation satisfied by $k_R$,
\begin{align}
\label{010316.3}
k_R^2+ \gamma^2 \cos^2(k'_R R)=0
\end{align}
 that results from Eq.~\eqref{290216.8} by requiring that $S(E_R)^{-1}=0$. For large $\gamma^2$, 
with $\gamma_0^2=[(n+1/2)\pi]^2$, Eq.~\eqref{010316.3}  can be expressed as
\begin{align}
k_R^2+R^{-2}{\cal O}(1)=0~,
\end{align}
where one has to notice that $k'_R R =\gamma_0\sqrt{1+(k_R R/\gamma_0)^2+(1+\varepsilon/\gamma_0)^2}$ and 
$\cos( (n+1/2) \pi+\delta)=-(-1)^n \delta+{\cal O}(\delta^3)$.
 
Let us also mention that the limit $ \gamma\to (n+1/2)\pi-0^+$ implies $a= - \infty$ (for fixed $n$), which is a 
virtual state at threshold. This is the ending (never reached) point that continuously connects with the resonance case that we 
are discussing by extrapolating the resonance mass below threshold.  
  The fact that $k_R$ does not scale with $\mu$ is an exemplification of the considerations that we already explored  
 within the ERE and heavy quark limit in  Sec.~\ref{sec:150216.1} (for $X\to 1$), cf. Eq.~\eqref{160216.1}.\footnote{Nevertheless, 
one has not to pursue too far the analogy with the spectroscopy of the square-well potential because this  depends on higher-order shape parameters 
in the ERE, and then it is more sensitive to the finer details of the toy model. In this respect we notice that
 for $|k_R|<1/R$ all the poles of $S(E)$ lie along the imaginary axis 
\cite{kolck.290216.1,nuss.010316.1}, while the ERE for this potential is valid  up to  $|k|\lesssim 1.3/R$ ($r/a=1$), as follows from the position of the 
closest zero of $t(E)$ at threshold. However, $a$ and $r$ parameterize the low-energy limit 
of the $T$-matrix and are sensitive to the global aspects of the potential, characterized by its strength and range.}

\section{Compositeness and width}
\label{sec:031115.2}

It is clear that  in the uncoupled ERE  with {\it real} values of $a$ and $r$  
the full widths of the $Z_b^{(\prime)}$ should correspond to the partial decay widths into the channels 
$B^{(*)}\overline B^{*}$. 
This is a bonus of the ERE approach employed here since in this way one does not need to apply a (heuristic) formula to reproduce 
 the width of a  resonance in terms of its coupling. This was e.g. the case  in Ref.~\cite{Oller1507} to account for the $\chi_{c1}p$
 width of $P_c(4450)$ (note the two-channel problem there, namely $J/\psi p$ and  $\chi_{c1}p$). However, the method of Ref.~\cite{Oller1507} 
allows one to drop several assumptions inherent to the ERE study of the uncoupled isovector $S$-wave $B^{(*)}\overline{B}^*$ scattering 
performed in the previous section.  This is convenient because  then 
one does not need to assume the particle content for the involved states nor 
scattering in just one partial wave (several partial waves per channel could be possible).   
In addition, this method is rather insensitive to the issue of the strength of nearby branch points from crossed-particle exchanges.\footnote{Nevertheless, 
it is assumed that a Lorentzian mass distribution with a constant coupling holds  for the resonance, which could be affected by 
the presence of nearby and  strong crossed-particle-exchange branch points.}

  We start by extending the brief discussion given in Ref.~\cite{Oller1507}, 
 which  establishes a rather clear picture for the width of the resonance in terms of $|\gamma|^2$ and the mass distribution induced by the 
finite width itself.
 On the one hand, we have the standard two-body decay formula \cite{pdg} of a resonance into one channel  in terms of its coupling squared. 
The corresponding equation in our normalization, Eq.~\eqref{271015.4}, can be easily deduced   by considering that $t(E)$ is saturated by 
the resonance contribution, $t(E)\to -\gamma^2/(s-s_R)$, and then requiring the fulfillment of the unitarity condition Eq.~\eqref{271015.4}. 
 This implies 
\begin{align}
\label{170316.2}
k=M_R\Gamma_R(E)/\gamma^2~.
\end{align} 
We then deduce the expression,
\begin{align} 
\Gamma^{(1)}=&\frac{k(M_R) |\gamma^2|}{ M_R}~.
\label{Gamma11}
\end{align}
  We take the modulus of the residue 
in Eq.~\eqref{Gamma11} because it is generally a complex number, cf. Eq.~\eqref{190216.9}. 
 The decay width formula Eq.~\eqref{Gamma11}  should be valid for a narrow resonance when the distances between 
$M_R$ and decay-channel thresholds are large  in comparison with the resonance width (this latter condition is not satisfied 
 by the $Z_b^{(\prime)}$ resonances.) 

Another point worth stressing with respect to Eq.~\eqref{Gamma11} is that this equation 
applies to the width of a resonance independently of its angular momentum because 
 the unitarity requirement Eq.~\eqref{271015.4} is valid for any partial wave (not only for $S$-wave). 
  The same can be said for the compositeness $X$ of Eq.~\eqref{190216.6}, that applies to any angular-momentum resonance, as it is 
clear from the analysis undergone in Ref.~\cite{Oller1508}. 
As a result, if a two-body particle decay channel appears in several partial waves for a given resonance (due to the mixing of 
orbital angular momentum and spin quantum numbers) the total decay width and  compositeness of the resonance 
 in such decay channel are given by Eqs.~\eqref{Gamma11} and \eqref{190216.6}, but with $|\gamma^2|$ corresponding 
to the sum of the residues of all partial waves involved, $|\gamma^2|=\sum_i |\gamma_i^2|$.  

  The expression  for $\Gamma^{(1)}$ 
 in terms of $\gamma_k^2$  follows from Eq.~\eqref{Gamma11} by taking into account the 
relationship between $\gamma^2$ and $\gamma_k^2$ given in Eq.~\eqref{190216.9}. The result is 
\begin{align}
\label{210216.1}
\Gamma^{(1)}=&\frac{2\gamma_k^2}{\mu} k(M_R)|k_R|~.
\end{align}
The more standard form of $\Gamma^{(1)}$, e.g. the one appearing in Ref.~\cite{pdg}, implies to use the resonance coupling
 squared $g^2$, which 
in terms of the residue $\gamma^2$ corresponds to
\begin{align}
\label{190216.12}
g^2=\left| \gamma^2 8\pi E_R\right|\simeq |\gamma^2| 8\pi M_R~. 
\end{align}
In the last step we have used  that  $M_R\gg \Gamma_R/2$ for the $Z_b^{(\prime)}$.  
In terms of $g^2$ Eq.~\eqref{Gamma11} becomes 
\begin{align} 
\Gamma^{(1)}=&\frac{k(M_R) g^2}{ 8\pi M_R^2}~.
\label{Gamma1}
\end{align}

On the other hand, we also have the more elaborated decay width formula \cite{NPA620,Oller1507}
\begin{align}
\label{Gamma2}
\Gamma^{(2)}=&\frac{g^2 }{16\pi^2}\int_{M_{\rm th}}^{+\infty}\frac{dW k(W)}{W^2}\frac{\Gamma_R}{(M_R-W)^2+\Gamma_R^2/4}\\
\label{Gamma2X}
=&\frac{X |k_R| M_R^2}{\pi \mu}\int_{M_{\rm th}}^{+\infty}\frac{dW k(W)}{W^2}\frac{\Gamma_R}{(M_R-W)^2+\Gamma_R^2/4}\,
\end{align}
where $k(W)$ is the three-momentum as a function of the total energy $W$, cf. Eq.~\eqref{271015.4k}. 
 The latter variable is  distributed around the resonance mass according to the Lorentzian mass distribution 
  of mass $M_R$ and width $\Gamma_R$. The last  Eq.~\eqref{Gamma2X} results from Eq.~\eqref{Gamma2} 
by taking into account the relationship between $g^2$ and $\gamma_k^2$, cf.  Eqs.~\eqref{190216.9}, \eqref{190216.6} and \eqref{190216.12}. 

We offer here  a brief derivation of Eq.~\eqref{Gamma2}. 
For that we allow to vary the invariant mass  $W$ of the two-body decay channel according to a mass distribution 
$F(W)$  around the nominal resonance mass $M_R$ and  denote by  $\mathcal{M}$ the corresponding resonance coupling/residue
 (which could also depend on $W$). 
In this way, instead of Eq.~\eqref{Gamma1} we have now the more general expression 
\begin{equation}
\label{dGamma2}
\Gamma=\int_{M_{\rm th}}^{+\infty} \frac{dW k(W)}{8\pi W^2}|\mathcal{M}|^2 F(M)~,
\end{equation}
where the lower limit of integration results because the two particles in the decay channels are the asymptotic ones.

For a narrow resonance one typically takes 
\begin{align}
\label{301015.2}
F(M)=& \delta(W-M_R)~,
\end{align} 
which, after being inserted in Eq.~\eqref{dGamma2},  leads to Eq.~\eqref{Gamma1}. 
When  finite-width effects are considered a standard and simple option is to use for $F(W)$ a Lorentzian mass distribution, 
\begin{align}
\label{301015.3}
F(W)=&\frac{1}{\pi}{\rm{Im}}\,\frac{1}{W-M_R-i\Gamma_R/2}\nonumber\\
=&\frac{1}{2\pi}\frac{\Gamma_R}{(W-M_R)^2+{\Gamma_R}^2/4}
\end{align}
and then Eq.~\eqref{Gamma2} results if additionally $|\mathcal{M}|^2\to g^2$, so that any possible energy dependence  is neglected.
  Indeed, Eq.~\eqref{Gamma1} stems also from Eq.~\eqref{Gamma2} in the limit of  zero width, being the later much more adequate 
to take into account finite-width effects. 
  For example,
 the mass of the $P_c(4450)$ \cite{LHCPenta} at its lower range within error bars lies indeed below the $\chi_{c1}p$ mass so that one cannot apply 
 Eq.~\eqref{Gamma1}, since then $k$ would become complex. 
 This caveat is solved with the use of Eq.~\eqref{Gamma2}, and  the error estimate of the calculated width can be performed rather straightforwardly
  from the experimental errors of the mass and width of $P_c(4450)$.

 For the practical use of  Eq.~\eqref{Gamma2} we distinguish between integrating up to $\infty$ or up to $M_R+n\Gamma_R$, with $n$ fixed such that 
$\Gamma^{(2)}$ coincides with the experimental value.  
 Proceeding in this way is justified because  the Lorentzian  mass distribution 
has a long tail which gives rise to a slow convergence of the integration in Eq.~\eqref{Gamma2}. 
 The resulting width from the integration up to $\infty$ is denoted by $\Gamma^{(2)}_>$ 
and the one up to $M_R+n\Gamma_R$ by $\Gamma^{(2)}_<$. 

We first consider the value for $g^2$ corresponding to the ERE results (last row of Table~\ref{tabar}).  
In this case,  consistency between ERE and the present method  requires that the calculated width
 in terms of the coupling be the same as the experimental one (the one imposed in the ERE analysis of Sec.~\ref{sec:031115.1}).  
 The numerical values of  $\Gamma^{(1)}$, $\Gamma^{(2)}_>$ and $\Gamma^{(2)}_<$ are  summarized in Table~\ref{tabXGamma}.  
 From the table, we see that these widths are compatible with the measured $\Gamma_R$ in Eq.~\eqref{mwzbl} within errors. 
The width $\Gamma^{(2)}_<$ reproduces exactly the experimental width with  $n\simeq 8$, that will be used in the following 
whenever we employ Eq.~\eqref{Gamma2} with a finite upper integration limit. 
 Of course,  $\Gamma^{(2)}_>$ is larger than $\Gamma^{(2)}_<$, as indicated by the subscripts, 
due to the larger integration region for a positive integrand in Eq.~\eqref{Gamma2}. 

\begin{table}[htbp]
\begin{center}
\begin{tabular*}{0.9\linewidth}{@{\extracolsep{\fill}}crr}
\hline\hline
 $\Gamma\,(\rm{MeV})$ &$Z_b(10610)$   &$Z_b(10650)$ \\
\hline
$\Gamma^{(1)}$            & $14.9 \pm 2.3$   & $9.5 \pm 2.1$  \\
$\Gamma^{(2)}_>$        & $21.9 \pm 3.3$   & $13.4 \pm 2.8$ \\
$\Gamma^{(2)}_<$        & $18.5 \pm 2.4$   & $11.3 \pm 2.1$   \\
\hline\hline
\end{tabular*}
\caption{Widths of the $Z_b^{(\prime)}$ obtained by using the coupling squared $g^2$ from the ERE, cf. Table~\ref{tabar}. 
For their calculation we have considered both $\Gamma^{(1)}$, Eq.~\eqref{Gamma1},  and $\Gamma^{(2)}$, \eqref{Gamma2}. 
For the latter case  the subscripts $>$ and $<$ refer to  the integration
 regions employed in Eq.~\eqref{Gamma2}, $[M_{\rm th}, \infty)$ and $[M_{\rm th}, M_R+n\Gamma_R]$, $n=8$, in order.  }
\label{tabXGamma}
\end{center}
\end{table}

\begin{table}[htbp]
\begin{center}
\begin{tabular*}{\linewidth}{@{\extracolsep{\fill}}lcc}
\hline\hline
 {} &$Z_b(10610)$   &$Z_b(10650)$ \\
\hline
$\Gamma=\Gamma_R$  & $X=0.76 \pm 0.12$  & $X=0.69 \pm 0.14$\\
$\Gamma=\Gamma_R \cdot \rm{Br}$  & $X_{\rm ex}=0.66 \pm 0.11$  & $X_{\rm ex}=0.51 \pm 0.10$\\
\hline\hline
\end{tabular*}
\caption{Compositeness coefficient $X$ in the limit case of saturating  the experimental total width $\Gamma=\Gamma_R$ (2nd row) and 
  $X_{\rm ex}$ (last row), that takes into account the  experimentally measured partial decay widths 
of $Z_b^{(\prime)}$ into  $B^{(*)}\overline B^*$ \cite{BelleZbconf2}.
}
\label{tab-discussion}
\end{center}
\end{table}

Now,  we follow  the  method of analysis of Ref.~\cite{Oller1507},  that heavily relies in the expressions for the width,
 Eq.~\eqref{Gamma2},  and compositeness $X$, Eq.~\eqref{190216.6}. 
 As indicated at the beginning of this Section, we do not need to assume in this way
 that the $Z_b^{(\prime)}$ resonances stem from uncoupled isovector $S$-wave 
 $B^{(*)}\overline{B}^*$ scattering, nor the applicability itself of the ERE up to the resonance pole positions, 
 as done in the ERE study of Sec.~\ref{sec:031115.1}. 

  Firstly, we perform two consistency checks between the previous ERE study in Sec.~\ref{sec:031115.1} and the new one based on the approach of Ref.~\cite{Oller1507}. 
On the one hand,  we take the same basic assumption as in the ERE, 
 so that the only relevant open channel around the resonance energy region
 would be the $B\overline{B}^*$($B^*\overline{B}^*$) for the $Z_b$($Z_b^\prime$) resonance. 
 As a result,  we assume  that $B^{(*)}\overline B^{*}$ saturate the experimental total decay  
widths of  $Z_b^{(\prime)}$, namely,  $\Gamma^{(2)}=\Gamma_R$. 
 From this requirement we can calculate  the  value of $X$ from Eq.~\eqref{Gamma2X}, with the resulting expression
\begin{align}
\label{300216.1}
X=&\frac{\mu \pi}{|k_R| M_R^2\int_{M_{\rm th}}^{W_+}\frac{dW k(W)/W^2}{(M_R-W)^2+{\Gamma_R}^2/4}}~,
\end{align} 
where $W_+=M_{\rm th}+n \Gamma^{(2)}$, and $n=8$ (as previously used to exactly reproduce the experimental width of the $Z_b^{(\prime)}$
  in the last column of Table~\ref{tabXGamma}). 
The calculated $X$ is given in the 2nd row of Table~\ref{tab-discussion} and perfectly agrees with the one found  
 in  the ERE study, cf. Table~\ref{tabar}. This
 consistency check would become a physical test of the ERE applicability in the study of the $Z_b^{(\prime)}$ resonances
 if $a$ and $r$ were actually measured experimentally in near threshold $1^{+}(1^{+})$ $B^{(*)}\overline{B}^*$ scattering.

For the second consistency check,  we assume that the two-body states $B^{(*)}\overline B^{*}$ saturate the components of $Z_b^{(\prime)}$, i.e., $X\to 1$. 
 This assumption also implies that the resonance should fully decay into $B^{(*)}\overline{B}^*$ because
 the coupling to any other open channel should tend to zero (otherwise $X$ would depart from 1). 
 In this case  $ \gamma_k^2$ takes its maximum possible value, $\gamma_k^2=1$. 
One can then deduce the
 value for the width  of the resonance to $B^{(*)}\overline{B}^*$ under such circumstances by making use of Eq.~\eqref{Gamma2X} 
with $X=1$ and  $\Gamma_R$ in the right-hand side  replaced by $\Gamma^{(2)}$. The following implicit equation for $\Gamma^{(2)}$ results:
\begin{align}
\label{210216.2}
1=&\frac{ |k(M_R-i\frac{\Gamma^{(2)}}{2})| M_R^2}{\pi \mu}\\
\times &\int_{M_{\rm th}}^{W_+}\frac{dW k(W)}{W^2}\frac{1}{(M_R-W)^2+{\Gamma^{(2)}}^2/4}~,\nn
\end{align}
with $ W_+$ evaluated with $n=8$ as   in Eq.~\eqref{300216.1}. One then obtains that $\Gamma^{(2)}$ becomes huge (in practical terms $\infty$) around several hundreds of MeV. 
This is in agreement with Eq.~\eqref{190216.11} for $X$ in the ERE, such that the limit $X\to 1$ with $M_R-M_{\rm th}\neq 0$  strictly requires that $\Gamma \to \infty$.

 Importantly,  the branching ratios for the partial decay widths of $Z_b^{(\prime)}$  into $B^{(*)}\overline B^{*}$  
 have been  measured  by Belle Collaboration in Ref.~\cite{BelleZbconf2},  cf. Eq.~\eqref{220216.3}. 
Although these figures are large, the $Br$'s are clearly less than 1, so that $\Gamma_R\neq \Gamma(Z_b^{(\prime)}\to B^{(*)}\overline{B}^*)$. 
 In order to take care of this 
experimental fact and provide a more accurate determination of $X$ one has to depart from the ERE study   
and use the method of Ref.~\cite{Oller1507}. 
The latter allows us to find  the actual compositeness coefficient $X$ (denoted by $X_{\rm ex}$)  by making use of 
 Eq.~\eqref{Gamma2X} with $\Gamma^{(2)}=\Gamma_R Br$. One then finds for $X_{\rm ex}$,
\begin{align}
X_{\rm ex}=&\frac{\pi \mu \,Br(Z_b^{(\prime)}\to B\overline{B}^*)}{|k_R|M_R^2\int_{M_{\rm th}}^{W_+}\frac{dW k(W)}{W^2}\frac{1}{(M_R-W)^2+\Gamma_R^2/4}}~.
\end{align}

 The results obtained with $W_+=M_{\rm th}+8\Gamma_R$  are shown in the last row of Table~\ref{tab-discussion}, 
being  almost identical to those that would result  if we had used instead $\Gamma^{(1)}$ from 
Eq.~\eqref{210216.1} (as we have checked).
The   $B\overline B^{*}$ weight calculated within the $Z_b$ is $(66 \pm 11)\%$, and the   $B^*\overline B^{*}$ one  for the $Z_b'$ 
  is $(51 \pm 10)\%$. We then conclude that the $B\overline{B}^*$ weight is  dominant  for the $Z_b$, while  the  $B^*\overline B^{*}$ component  is
around  one half of the total composition of the $Z_b'$. As a result, other non $B^{(*)}\overline{B}^*$ 
components play a more important role for the latter, being appreciable  for both resonances. 
This is also clear by comparing $X_{\rm ex}$ in Table~\ref{tab-discussion} with the compositeness coefficient deduced in the ERE study, cf. 3rd row of Table~\ref{tabar}. 
For the $Z_b$ state both calculations have similar central values while for the $Z_b'$ they are more different, though still compatible within error bars. 
 This latter fact also reflects the smaller branching ratio  for the $Z_b'$ in Eq.~\eqref{220216.3}. 
 We also  mention that our results for $X_{\rm ex}$  are similar to  those in  Ref.~\cite{ChenGY}, 
where $X$  for each resonance is estimated to be around 60\%.

\section{ERE vs. Breit-Wigner and Flatt\'e parameterizations}
\label{sec.170316.1}

We now address   the question about the applicability 
of Breit-Wigner functions in the experimental analyses of Refs.~\cite{BelleZb,BelleZbconf2,BelleZbconf3} 
for the $Z_b^{(\prime)}$ due to the closeness of the thresholds of the $B^{(*)}\overline{B}^*$ states, 
 an issue  stressed e.g. in Refs.~\cite{chen.100316.1,Cleven1,Hanhart.110316.1}.
 The ERE is  adequate to check whether a Breit-Wigner (or Flatt\'e) function is appropriate  since the applicability of ERE is not affected by 
the threshold singularity associated with the $B^{(*)}\overline{B}^*$.
 For definiteness let us take the 
central values of $a$ and $r$ in Table~\ref{tabar} for the $Z_b$ (analogous conclusions would equally apply to the $Z_b'$). 
We are going to compare $|t(E)|^2$ calculated within the ERE, cf. Eq.~\eqref{271015.1}, with i) a Breit-Wigner function of constant width, $bw(E)$, 
\begin{align}
\label{160311.1}
bw(E)=&|t(M_R)|^2 \frac{\Gamma_R^2/4}{(E-M_R)^2+\Gamma_R^2/4}
\end{align}
and ii) with a Flatt\'e parameterization \cite{Flatte.160311.1}, $ft(E)$, 
\begin{align}
\label{160311.2}
ft(E)=&\frac{f_0}{|E-M_R'+i \,\Gamma_R(E)/2|^2}~.
\end{align}
 Here $\Gamma_R(E)$ is the analytically-continued energy-dependent width that in the 1st RS along the real axis corresponds to, 
cf. Eq.~\eqref{170316.2},  
\begin{align}
\label{160311.3}
\Gamma_R(E)=&\frac{2\gamma_k^2}{\mu} k(E)|k_R|~,~E\geq M_R~,\nn\\
=&i \frac{2\gamma_k^2}{\mu} |k(E) k_R|~,~E<M_R~.
\end{align}
 In addition,  $M'_R$ is fixed such that the real part of the complex number in the denominator 
of Eq.~\eqref{160311.2} vanishes at the pole position $E=E_R$ (in the 2nd RS). This implies the equation 
\begin{align}
\label{170316.1}
M'_R=M_R+\gamma_k^2k_i |k_R|/\mu~.
\end{align}
The normalization factors $|t(M_R)|^2$ in Eq.~\eqref{160311.1} and $f_0$ in Eq.~\eqref{160311.2} are fixed such that the maximum value of 
 the Breit-Wigner and Flatt\'e parameterizations coincides with the maximum of $|t(E)|^2$.

\begin{figure}
\begin{center}
\includegraphics[width=0.45\textwidth]{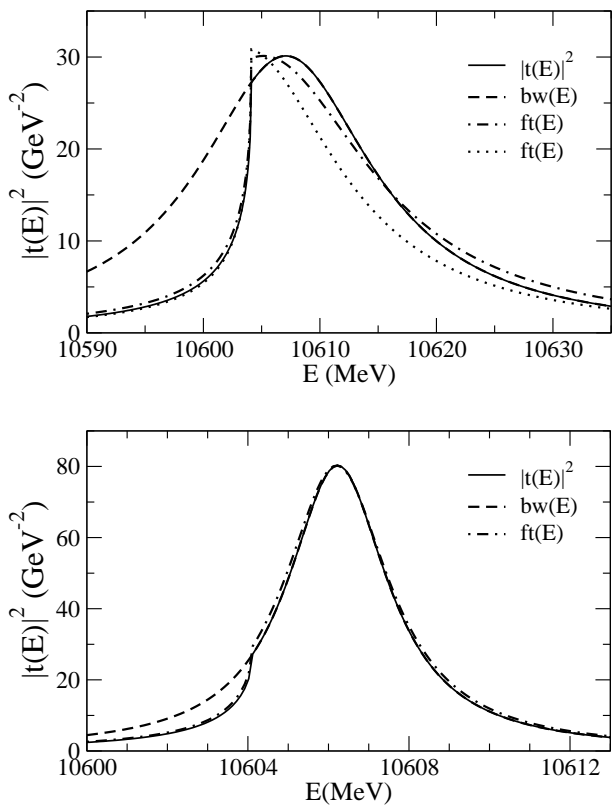}  
\end{center}
\caption[pilf]{\protect {Amplitude squared for uncoupled isovector $S$-wave $B\overline{B}^*$ 
scattering: $|t(E)|^2$ with $t(E)$ given by the ERE (solid line), Eq.~\eqref{271015.1};  
the Breit Wigner function $bw(E)$ (dashed line), Eq.~\eqref{160311.1};  
the Flatt\'e parameterization $ft(E)$ (dash-dotted line), Eq.~\eqref{160311.2}. The dotted line in the top panel 
corresponds to a Flatt\'e parameterization in which $M'_R=M_{\rm th}+\frac{|k_R|^2}{2\mu}$, cf. Eq.~\eqref{120316.2}. 
\label{fig.160311.1}
}}
\end{figure}

We perform this comparison in the top panel of Fig.~\ref{fig.160311.1} where the solid line corresponds to $|t(E)|^2$, the dashed curve to 
$bw(E)$ and the dash-dotted one to $ft(E)$. On the one hand, we see that $bw(E)$ lies on top of $|t(E)|^2$ above threshold, while 
 below it they do not follow each other due to the cusp effect in $|t(E)|^2$. This  simply reflects the fact that the pole position of  
the $Z_b$ lies on the 2nd RS, and this RS connects continuously with the physical energy axis only above the $B\overline{B}^*$ 
threshold.\footnote{The overlapping between different RS's is discussed in detail in Ref.~\cite{oller.130316.1}.}
 Below this threshold  the form of the rise is a typical cusp effect which is enhanced because the pole of the $Z_b$ fixes the size of $t(E)$
 at the threshold point. On the other hand, we observe that the Flatt\'e parameterization $ft(E)$ can  reproduce accurately the 
cusp effect for $E<M_{\rm th}$, while it agrees worse with $|t(E)|^2$ than the Breit-Wigner function for $E>M_{\rm th}$. 
 The Flatt\'e parameterization as a function of energy shows above threshold  a 
different qualitative form compared with $|t(E)|^2$, although they are not very different quantitatively. 

 The reason for this  disagreement above threshold lies in the fact that  $\Gamma_R$ is significantly 
larger than $M_R-M_{\rm th}$ and this softens the energy dependence of $|t(E)|^2$  around the $B\overline{B}^*$ threshold 
compared with the abrupter effects implied by Eq.~\eqref{160311.3} when implemented in Eq.~\eqref{160311.2}.\footnote{Notice
 that in the Laurent series around the pole position of the 
resonance the function $t(E)$ and its derivatives are evaluated at $E_R=M_R-i\Gamma_R/2$. This is why for 
 $\Gamma_R\gg |M_R-M_{\rm th}|$  the effects of the branch point for $M_R\simeq M_{\rm th}$ are softened.}  
 According to the previous discussion, in order to end with a situation in which the Flatt\'e parameterization matches  $|t(E)|^2$ above threshold,
 we  decrease the width of the pole obtained from the ERE approximation  by decreasing $r$ and keeping $a$ fixed (so that $X$ decreases). 
If we take $r=-5.5$~fm the pole position is now $(10606.2-i 3.0)$~MeV (with $X=0.32$); the width is six times smaller, while the resonance 
mass changes very little. Since now the imaginary part of the pole position has a size comparable with $M_R-M_{\rm th}\simeq 2$~MeV 
we should then expect a better agreement with the Flatt\'e distribution. 
This is indeed what happens as plotted in the bottom panel of Fig.~\ref{fig.160311.1}, where the Flatt\'e
 distribution is able to reproduce faithfully $|t(E)|^2$ in the whole energy range shown. Once more, 
the Breit-Wigner parameterization with constant width perfectly matches $|t(E)|^2$  above threshold.

We can indeed rewrite Eq.~\eqref{271015.1} by taking into account the presence of the resonance, so that 
Eq.~\eqref{170216.3} is fulfilled. Equation \eqref{271015.1} then becomes
\begin{align}
\label{120316.1}
t(E)=&\frac{1}{(\frac{r}{2}(k-k_R)-i)(k+k_R)}\\
\label{120316.2}
=&\frac{-k_i/\mu}{E-M_{\rm th}-\frac{|k_R|^2}{2\mu}+i k k_i/\mu}~.
\end{align}
The difference with respect to the Flatt\'e parameterization in Eq.~\eqref{160311.2} stems from the fact that now
  one has 
\begin{align}
\label{120316.3}
\gamma(E)=-\frac{2\gamma_k^2 k k_r}{\mu}
\end{align}
 instead of $\Gamma_R(E)$, with $|k_R|$ substituted by $-k_r$. Again in the limit of small width both expressions agree 
but for $\Gamma_R\gg |M_R-M_{\rm th}|$ they are markedly different, because in such a case 
$|k_R|\simeq -k_r \sqrt{2}$, cf. Eq.~\eqref{150216.4}, and then  $\Gamma_R(E)\simeq \gamma(E)\sqrt{2}$. 
 Due also to this  reason, if we directly identify $M_R'$  with $M_{\rm th}+\frac{|k_R|^2}{2\mu}$ in $ft(E)$, as read from Eq.~\eqref{120316.2}, 
 we  obtain different results, corresponding to the dotted line in the top panel of Fig.~\ref{fig.160311.1}. The disagreement between $ft(E)$ and  $|t(E)|^2$ 
in this case is quite worse than the one between the solid and dash-dotted lines in
 the top panel of Fig.~\ref{fig.160311.1}.  The adjusting process for $M_R'$, 
Eq.~\eqref{170316.1}, reduces the differences. In conclusion, for the case of the $Z_b^{(\prime)}$ resonances 
the use of Breit-Wigner functions to fit data \cite{BelleZb,BelleZbconf2,BelleZbconf3} seems to be
  a reasonable approach, as long as the resonance mass is clearly above threshold. This conclusion is reflected by  
similarity between the solid and dashed lines in  Fig.~\ref{fig.160311.1}. Indeed, the peak position 
 fixes the Breit-Wigner mass in $bw(E)$ in a precise way (in agreement with $M_R$), 
while one could certainly obtain a reasonable estimate  for the width 
 by comparing  $|t(E)|^2$ and the approximate $bw(E)$ curve. Nonetheless, sounder theoretical 
 parameterizations of data should be pursued so as to pin down resonance parameters more accurately. 
In this respect, the modified version of the Flatt\'e parameterization corresponding to 
Eq.~\eqref{120316.2}, exactly reproduces $|t(E)|^2$ from the ERE.


\section{Summary and conclusions}
\label{sec:conclusion}

In this work we study the resonances $Z_b(10610)$ ($Z_b$) and $Z_b(10650)$ ($Z_b'$)  
based on analyticity and unitarity as basic principles of scattering theory. 
The dynamics is encoded in the pole positions of the resonances, which 
allows us to determine from unitarity the couplings of these resonances to $B\overline{B}^*$ and $B^*\overline{B}^*$, in order. 
These basic principles also determine  the compositeness coefficient ($X$)
of a resonance  into an open two-body channel, as shown in Ref.~\cite{Oller1508}. 
We adapt these results to nonrelativistic scattering and obtain that $X$ can be 
calculated for resonances whose mass is larger than threshold. 
 By employing the  uncoupled effective range expansion (ERE), up to  including ${\cal O}(k^2)$ terms, 
  $X$  can be directly expressed in terms of the measurable 
scattering length ($a$) and effective range ($r$) as $X=1/\sqrt{2r/a-1}$. 
 With $k_R=k_r+ik_i$ ($k_i>0, k_r<0$), the two-body three-momentum at the resonance pole position, one can also express $X$ in 
terms of the resonance mass $M_R$ and width $\Gamma_R$ as $X=-k_i/k_r$. 
 For a small ratio $2|M_R- M_{\rm th}|/\Gamma_R$ this expression simplifies to 
 $X=1-2|M_R- M_{\rm th}|/\Gamma_R+{\cal O}((|M_R- M_{\rm th}|/\Gamma_R)^2)$, being $M_{\rm th}$  the threshold energy.

We first perform an ERE study of  uncoupled  isovector  $S$-wave $B^{(*)}\overline B^{*}$ scattering,
  because it has  the same  quantum numbers as $Z_b^{(\prime)}$ \cite{BelleZb}
 and these resonances lie in the vicinity of the open-bottom meson thresholds. 
It is then worth first focusing just on $S$-wave. Nevertheless, the 
associated three-momentum at the resonance pole position ($k_R$) has a 
modulus around $m_\pi$, with $m_\pi$ the pion mass. 
In connection with this, we  argue that although the branch point
 due to one pion exchange (at $k=i m_\pi/2$) could severely restrict
 the applicability of ERE to $B^{(*)}\overline{B}^*$ scattering up to the resonance pole positions,
 this seems not finally be the case  because of the perturbative nature of pions in the present systems. 
Namely, we consider estimates based on power counting in effective field theory \cite{pavon.pwc.230216.1} which shows that 
the iteration of pion exchanges is suppressed by a large expansion scale $\Lambda$  for 
the isovector $B^{(*)}\overline{B}^*$ systems ($\Lambda \gg m_\pi$). This suppression is in  line with the observation  of Ref.~\cite{Oset}, 
 also reflected in specific calculations \cite{Oset,XLiu2011,Ke.170316.1},  that the  exchanges of  light $q\bar{q}$ mesons are OZI suppressed for the isovector $B^{(*)}\overline{B}^*$ scattering. 
 The very same power counting study of Ref.~\cite{pavon.pwc.230216.1} also establishes that coupling effects between 
 different channels should be suppressed, at least a next-to-leading effect. This also matches with the experimental 
 large  branching ratios of the $Z_b^{(\prime)}$ into $B^{(*)}\overline{B}^*$, measured in Ref.~\cite{BelleZbconf2}. 
As a result, the  ERE study of the uncoupled isovector $S$-wave $B^{(*)}\overline{B}^*$ scattering seems an adequate first  approximation 
to the problem. 

 Within  this working assumption, we apply the ERE up to including ${\cal O}(k^2)$ and the  
real parameters involved, $a$ and $r$, are fixed  so as to reproduce the pole positions of the 
$Z_b^{(\prime)}$ resonances according to their masses and widths. 
The resulting values of $a$ and $r\sim -(1-2)$~fm for both resonances have a size of around the standard range of strong interactions. 
This in turn indicates that the  $B^{(*)}\overline{B}^*$ scattering near threshold would rather correspond to a potential scattering problem, which also  
implies that other compact components in the $Z_b^{(\prime)}$ states would  play a relatively minor role. 
This is based on the observation that otherwise the nearby presence of a CDD pole would  
imply large contributions to $|r|$, cf. Eq.~\eqref{281015.5b}.
 Note that  a priori, within an uncoupled $S$-wave ERE for $B^{(*)}\overline B^{*}$ scattering, the magnitude  of the effective range $r$ could be very different
  compared to the standard potential scattering problem. 
  See Ref.~\cite{guo16} for a concrete example of a physical system with a resonance on top of thresholds 
 that requires large values  for $|r|$ and small ones of $X$.

 This conclusion on the inner structure of the  $Z_b^{(\prime)}$ states is further supported through the 
quantitative calculation of the compositeness $X$ within ERE. We find that $X=1/\sqrt{2r/a-1}$ and obtain the 
numerical values  $X=0.75\pm 0.15$ and $0.67\pm 0.16$ for the $Z_b$ and $Z_b'$, in order.
 These numbers are rather close to 1, which occurs whenever $ \Gamma_R\gg M_R-M_{\rm th}$ ($M_R\geq M_{\rm th})$. This is a qualitative feature 
realized in many $XYZ$ resonances and that indicates that the corresponding states would have an important two-heavy-particle component in their compositeness (as long as 
the application of uncoupled ERE is justified).  
 We also derive a scaling law for the width of purely composite two-heavy-particle
  resonances in $S$-wave such that $\Gamma\sim 2\Lambda_{\rm QCD}^2/\mu\gg M_R-M_{\rm th}$, where $\Lambda_{\rm QCD}\simeq 200$~MeV, 
 $\mu$ is the reduced mass of the two-heavy particles.  

In the last part of the paper, we apply the method of Ref.~\cite{Oller1507} that only relies on the expressions for compositeness and the partial 
decay width of a resonance (taking  into account its mass distribution). In particular, its applicability is
 not directly affected neither by  nearby branch points from crossed-particle exchanges nor by the presence of other channels. Although, in connection with this,
 the validity of the Lorentzian mass distribution for the resonance is assumed to hold.  
 In this way we can  take into account the fact that the branching ratios
 of $Z_b^{(\prime)}$ to $B^{(*)}\overline{B}^*$, although rather large, are less than 1. 
 After showing that consistent  results with the ERE study are obtained
 under the same basic assumption (that the total width of the resonance is saturated by its partial decay width into $B^{(*)}\overline{B}^*$), 
 we apply the method of Ref.~\cite{Oller1507}  to find the actual value of compositeness ($X_{\rm ex}$). 
  We finally obtain that $X_{\rm ex}=0.66\pm 0.11$ and $X_{\rm ex}=0.51\pm 0.10$ for the $Z_b$ and $Z_b'$, 
respectively, in accordance with the experimentally measured branching ratios. The  $B^{(*)}\overline{B}^*$ components
 found indicate that the $B\overline{B}^*$ is dominant among the components of the $Z_b$, but around half of the composition for the $Z_b'$. 
Other contributions then also play an important role in the inner composition of these resonances (particularly for the latter). 

We also perform a comparison between the ERE approximation for the isovector $S$-wave $B^{(*)}\overline{B}^*$ scattering amplitude squared,
$|t(E)|^2$, and the Breit-Wigner and Flatt\'e parameterizations. This comparison could  shed light on the issue of whether it 
is appropriate to apply Breit-Wigner functions to fit data in the experimental study of the $Z_b^{(\prime)}$ resonances.  
 The reason is because the applicability of the ERE is not affected by the nearby $B^{(*)}\overline{B}^*$ threshold,  
   while this is the case for both Breit-Wigner and Flatt\'e parameterizations.  
 We obtain that there is a marked cusp effect for energies below threshold that can be well reproduced by a Flatt\'e 
parameterization. However,  the use of the Breit-Wigner function with constant width for resonance peaks above threshold accurately reproduces 
$|t(E)|^2$ for $E>M_{\rm th}$ (including the maximum of the amplitude squared which fixes the resonance mass). We have also derived 
a variant of Flatt\'e parameterization that exactly reproduces $|t(E)|^2$ from ERE. All in all,  
 we conclude that it seems  plausible to use Breit-Wigner functions to study the $Z_b^{(\prime)}$, though sounder 
  parameterizations are clearly required to improve accuracy.

 Needless to say that the method developed in Refs.~\cite{Oller1507,guo16} and in this work, employed here to  study  the $Z_b^{(\prime)}$ resonances, 
 could be also applied to other near-threshold resonances, and work along this direction is ongoing.

\section*{Acknowledgements}
J.A.O. would like to thank David Rodr\'{\i}guez Entem for discussions on the sign of the ERE parameters and local interactions.  
This work is supported in part by the MINECO (Spain) and ERDF (European Commission) grant FPA2013-40483-P and the Spanish
 Excellence Network on Hadronic Physics FIS2014-57026-REDT,  the National Natural Science Foundation of China (NSFC) 
under Grant Nos.~11575052 and 11105038, the Natural Science Foundation of Hebei Province with contract No.~A2015205205,
the grants from the Education Department of Hebei Province under contract No.~YQ2014034,
the grants from the Department of Human Resources and Social Security of Hebei Province with contract No.~C201400323, 
the Sino-German Collaborative Research Center ``Symmetries and the Emergence of Structure in QCD'' (CRC~110) 
co-funded by the DFG and the NSFC.


\end{document}